\documentclass[twoside,twocolumn,9pt,linenumbers]{article}
\usepackage{extsizes}
\usepackage[numbers,sort&compress,square]{natbib}
\usepackage[version=3]{mhchem}
\usepackage[left=1.5cm, right=1.5cm, top=1.8cm, bottom=2.5cm]{geometry}
\usepackage{balance}
\usepackage{times,mathptmx}
\usepackage{graphicx} 
\usepackage{lastpage}
\usepackage{changepage}
\usepackage[format=plain,justification=justified,singlelinecheck=false,font={stretch=1.125,small},labelfont=bf,labelsep=space]{caption}
\usepackage{float}
\usepackage{fancyhdr}
\usepackage{fnpos}
\usepackage[english]{babel}
\usepackage{mathtools}
\usepackage{array}
\usepackage{droidsans}
\usepackage{charter}
\usepackage[T1]{fontenc}
\usepackage[usenames,dvipsnames]{xcolor}
\usepackage{setspace}
\usepackage[compact]{titlesec}
\usepackage{hyperref}

\hypersetup{
    colorlinks,
    linkcolor={blue},
    citecolor={blue},
    urlcolor={blue!80!black}
}
\usepackage{xr}
\usepackage{amsmath}
\usepackage{amssymb}
\usepackage{color}
\usepackage[english]{babel}
\usepackage{enumerate}
\usepackage{float}
\usepackage{placeins}
\usepackage{graphicx}
\usepackage{bm}
\usepackage{mhchem}

\usepackage{relsize} 

\usepackage{soul} 
\usepackage[switch]{lineno}

\newcommand*{\myeqref}[2][Eq.~]{#1(\ref{#2})}
\newcommand*{\myfigref}[2][Fig.~]{#1\ref{#2}}
\newcommand*{\mysecref}[2][Sec.~]{#1\ref{#2}}
\newcommand{\supplref}[2][]{#1\ref{#2} of the SI}
\newcommand{\supplsecref}[2][Sec.~]{#1\ref{#2}  of the SI}
\newcommand{\supplfigref}[2][Fig.~]{#1\ref{#2} of the SI}

\newcommand\blfootnote[1]{%
  \begingroup
  \renewcommand\thefootnote{}\footnote{#1}%
  \addtocounter{footnote}{-1}%
  \endgroup
}
\begin{document}

\setlength\bibsep{1pt}

\setlength{\columnsep}{6.5mm}
\makeatletter 
\newlength{\figrulesep} 
\setlength{\figrulesep}{0.5\textfloatsep}

\makeatother

\twocolumn[
  \begin{@twocolumnfalse}
  
{\noindent \Huge The solid-state Li-ion conductor Li$_\textbf{7}$TaO$_\textbf{6}$: A combined computational and experimental study}

\vspace*{0.3cm}

{\noindent \LARGE  \large{Leonid Kahle $^{\text{a},*,\ddag}$,
 Xi Cheng $^{\text{c},\ddag}$,
 Tobias Binninger $^{\text{b},\ddag}$,
 Steven David Lacey $^\text{d}$,
 Aris Marcolongo $^\text{b}$,
 Federico Zipoli $^\text{b}$,
 Elisa Gilardi $^\text{c}$,
 Claire Villevieille $^\text{d}$,
 Mario El Kazzi $^\text{d}$,
 Nicola Marzari $^\text{a}$,
 and Daniele Pergolesi $^{\text{c,d},*}$
 }

}


{\small \noindent \textit{$^\text{a}$~Theory and Simulation of Materials (THEOS), and National Centre for Computational Design and Discovery of Novel Materials (MARVEL),
 \'{E}cole Polytechnique F\'{e}d\'{e}rale de Lausanne, CH-1015 Lausanne, Switzerland}\\
\textit{$^\text{b}$~IBM Research--Zurich, CH-8803 R\"uschlikon, Switzerland}\\
\textit{$^\text{c}$~Laboratory for Multiscale Materials Experiments,
Paul Scherrer Institut, CH-5232 Villigen PSI, Switzerland}\\
\textit{$^\text{d}$~Electrochemistry Laboratory, Paul Scherrer Institut, CH-5232 Villigen PSI, Switzerland}\\
}

\hrulefill

\vspace*{0.3cm}

\begin{tabular}{m{2.5cm} p{13cm} }

& {\large A B S T R A C T} \\
& \hrulefill \\
& 
\noindent\normalsize{
We study the oxo-hexametallate Li$_7$TaO$_6$ with  first-principles and classical molecular dynamics simulations, obtaining a low activation barrier for diffusion of $\sim$0.29~eV and a high ionic conductivity of $5.7 \times 10^{-4}$~S\,cm$^{-1}$ at room temperature (300~K).
We find evidence for a wide electrochemical stability window from both calculations and experiments, suggesting its viable use as a solid-state electrolyte in next-generation solid-state Li-ion batteries.
To assess its applicability in an electrochemical energy storage system, we performed electrochemical impedance spectroscopy measurements on multicrystalline pellets, finding substantial ionic conductivity, if below the  values predicted from simulation.
We further elucidate the relationship between synthesis conditions and the observed ionic conductivity using X-ray diffraction,  inductively coupled plasma optical emission spectrometry, and X-ray photoelectron spectroscopy, and study the effects of Zr and Mo doping.
}\\
& \hrulefill \\

\end{tabular}

 \end{@twocolumnfalse} \vspace{0.6cm}

\vspace*{0.6cm}

  ]

\blfootnote{
$^*$ Corresponding authors. \\
\hspace*{0.64cm} \textit{E-mail addresses}: leonid.kahle@epfl.ch, daniele.pergolesi@psi.ch
}
\blfootnote{\ddag~These authors contributed equally to this work}

\renewcommand*\rmdefault{bch}\normalfont\upshape
\rmfamily
\section*{}

\renewcommand{\textfraction}{0.02}
\renewcommand{\floatpagefraction}{0.1}
\renewcommand{\bottomfraction}{1.0}
\renewcommand{\topfraction}{1.0}
\setcounter{topnumber}{1}
\setcounter{bottomnumber}{1}
\setcounter{totalnumber}{2}
\setstretch{1.125}
\vspace*{-4ex}

\section{Introduction}
\label{sec:introduction}

Li-ion batteries power a critical set of portable technologies~\cite{armand_building_2008} and are key to the deployment of electric vehicles, necessary to mitigate the carbon footprint of the vehicle fleet.
It is important to overcome the constraints on safety~\cite{quartarone_electrolytes_2011, balakrishnan_safety_2006} and power/energy density~\cite{schaefer_electrolytes_2012} of today's Li-ion batteries, largely due to the use of liquid and organic electrolytes.
Solid-state electrolytes (SSE) are a promising alternative for next-generation batteries~\cite{li_solid_2015, kato_high-power_2016, manthiram_lithium_2017} and are being intensely researched using experiments and simulations~\cite{knauth_inorganic_2009,bachman_inorganic_2016, ceder_predictive_2018}.
Low electronic mobility, a large electrochemical stability window, good mechanical stability, and high Li-ionic conductivity are key properties that must be satisfied by any solid-state ionic conductor to qualify for potential applications as a SSE~\cite{thangadurai_recent_2006,manthiram_lithium_2017}.

Several structural families  have been and are being investigated as candidates for SSE application.
The Li-conducting garnets, with the well-known representatives  \ce{Li5La3Ta2O12} and \ce{Li7La3Zr2O12}~\cite{thangadurai_novel_2003,murugan_fast_2007,ferraresi_electrochemical_2018}, are one example of a thoroughly studied family of Li-ionic conductors.
Aliovalent substitutions on the Li, La, and Zr/Ta sites have led to a large variety of related structures~\cite{thangadurai_garnet-type_2014,thangadurai_fast_2015},
aiding in the stabilization of the ionically faster conducting cubic phase, and introducing vacancies to facilitate Li-ion diffusion.
Recent work focuses on characterizing dendritic growth through the garnet-based SSE~\cite{sudo_interface_2014},
and the interface between the electrolyte and electrode~\cite{broek_interface-engineered_2016,han_interphase_2018}.
Li-superionic conductors (LISICON) comprise another family of compounds explored for high ionic conductivity.
One such compound is Li$_{3+x}$(P$_{1-x}$Si$_x$)O$_4$, a solid solution of Li$_3$PO$_4$ and Li$_4$SiO$_4$~\cite{hu_ionic_1977,deng_structural_2015};
Substitutions of P and Si with B, Al, Zr, Ge, Ti, or As led to the discovery of several fast conductors~\cite{shannon_new_1977,rodger_li+_1985,deng_enhancing_2017},
and the substitution of sulfur with oxygen resulted in the sub-family of thio-LISICONs~\cite{kanno_synthesis_2000,kanno_lithium_2001,murayama_structure_2002,murayama_synthesis_2002,murayama_material_2004,wu_surface_2017}.
The increased ionic conductivity of these compounds compared to the respective oxygen-based LISICONs is attributed to a higher polarizability of the anion~\cite{bachman_inorganic_2016, kanno_lithium_2001}.
One of the best ionic conductors,  tetragonal \ce{Li10GeP2S12} (LGPS)~\cite{kamaya_lithium_2011},  with an ionic conductivity of 12~mS\,cm$^{-1}$ at room temperature, is found in this family.
However, sulfur substitution has a deleterious effects on the electrochemical stability~\cite{zhu_origin_2015} of the SSE, compared to their oxygen counterparts.
In addition, oxides have higher bulk and shear moduli than sulfides, and this increased mechanical stability could benefit the suppression of Li-metal dendrite growth~\cite{tikekar_design_2016}.

Switching to  thin-film batteries could also lead to technological breakthroughs due to beneficial mechanical properties (lower susceptibility to volume changes) and low resistance due to reduced dimensions~\cite{bates_fabrication_1993,quartarone_electrolytes_2011,kim_review_2015}.
Lithium phosphorus oxynitrides (LiPON) Li$_x$PO$_y$N$_z$ ($x=2y+3z-5$) can be grown into amorphous thin films, but its activation barriers at $\sim$0.55~eV, and ionic conductivity at room temperature of $2.3\times 10^{-6}~\mathrm{S\,cm^{-1}}$ are significantly worse than LGPS~\cite{yu_stable_1997}.
Still, these shortcomings are  compensated by reductions in the electrolyte thickness for use in thin-film batteries. 
Growing thio-LISICONs or garnets as thin films has also been investigated recently~\cite{wang_highly_2013,rawlence_chemical_2016,pfenninger_low_2019}.
Still, the search for more candidate SSE for Li-ion batteries is of major importance~\cite{avdeev_screening_2012,d.sendek_holistic_2017,muy_high-throughput_2019}, and novel discoveries could lead to rapid advances in the field.

\begin{figure}[t]
    \centering
    \includegraphics[width=0.8\hsize]{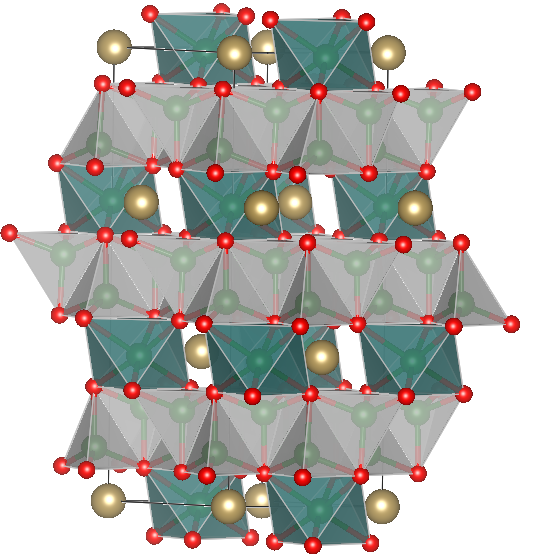}
    \caption{The structure of \ce{Li7TaO6}:
    Tantalum, lithium, and oxygen atoms are shown as gold, green, and orange spheres, respectively.
    The coordination of Li sites is highlighted by polyhedra.
    Octahedrally-coordinated sites (green polyhedra) are within the plane of Ta atoms, and tetrahedrally-coordinated sites (grey polyhedra) are in between planes of Ta atoms. 
    }
    \label{fig:tantalate}
\end{figure}

In this work, we highlight  the oxo-hexametallate \ce{Li7TaO6} and study its ionic conductivity, electrochemical stability, and applicability as SSE,  using both atomistic simulations and experimental characterization. 
This material has been characterized experimentally before, first by Scholder and Gl\"aser~\cite{scholder_uber_1958,scholder_uber_1964}, and later by Wehrum and Hoppe~\cite{wehrum_zur_1994}, and its structure is shown in \myfigref{fig:tantalate}.
Synthesis of a single-crystalline sample is mentioned in the latter reference, where the authors mixed Li$_2$O and Ta$_2$O$_5$ in a 7.7:1 molar ratio and annealed the mixture for 156 days at 1000~$^\circ$C.
While the aforementioned studies only refer to synthesis and structural properties, a few studies also report measurements of the Li-ionic conductivity in \ce{Li7TaO6}.
Delmas \textit{et al.}~\cite{delmas_conducteurs_1979} studied the phases \ce{Li8MO6} (M = Zr, Sn) and \ce{Li7LO6} (L = Nb, Ta) in 1979, finding the Li-tantalate to have the highest ionic conductivity among the four structures, namely $4.3 \times 10^{-8}~\mathrm{S\,cm^{-1}}$ at 300~K, and an activation barrier of 0.66~eV, measured with impedance spectroscopy on multicrystalline samples of 80\% density with respect to the theoretical density.
The authors suggested for this structure a two-dimensional diffusive pathway within the Ta layers.
In 1984, Nomura and Greenblatt~\cite{nomura_ionic_1984} studied 
\ce{Li7TaO6},
measuring ionic conductivities as low as of $3.7 \times 10^{-8}~\mathrm{S\,cm^{-1}}$  for at room temperature, and an activation barrier of 0.46~eV in a low temperature regime, and 0.67~eV in a high temperature regime, with the transition occurring at approximately $50~^\circ$C.
The authors managed to increase the ionic conductivity of \ce{Li7TaO6} by doping it with Nb, Bi, Zr, or Ca on the Ta site, with the highest ionic conductivity ($3.4 \times 10^{-7}~\mathrm{S\,cm^{-1}}$) reached for Li$_{7.4}$Ta$_{0.6}$Zr$_{0.4}$O$_6$.
In the most recent work on \ce{Li7TaO6}, M\"uhle \textit{et al.}~\cite{muhle_new_2004}  determined in 2004 its ionic conductivity using impedance spectroscopy, reporting a value of $1.53 \times 10^{-7}~\mathrm{S\,cm^{-1}}$ at 50~$^\circ$C and an activation barrier of 0.29~eV at 400--700~$^\circ$C, and 0.68~eV at 50--400~$^\circ$C.

In summary, we note that only three studies have -- to our knowledge -- investigated ionic diffusion in \ce{Li7TaO6} in the last four decades, with reasonable agreement in the Arrhenius behavior of the ionic conductivity, with all studies reporting values between 0.66~eV and 0.68~eV.
However, a more diffusive regime with a lower barrier is found above 400~$^\circ$C degrees by M\"{u}hle \textit{et al.}~\cite{muhle_new_2004}, with no other work having studied this regime.
Nomura and Greenblatt report a lower barrier at lower temperature~\cite{nomura_ionic_1984}.
Interestingly, this is not confirmed by the other studies, even though this regime was probed by Delmas \textit{et al.}~\cite{delmas_conducteurs_1979} and M\"uhle \textit{et al.}~\cite{muhle_new_2004}.
In addition, no prior work reports the electrochemical stability window of \ce{Li7TaO6}, and no atomistic simulation were performed on this material to elucidate the ionic transport mechanism.

The structure, deposited into the  Inorganic Crystal Structure Database~\cite{belsky_new_2002}, was identified by us as a fast ionic conductor during a computational screening for new solid-state electrolytes employing the pinball model~\cite{kahle_modeling_2018,kahle_high-throughput_2019}, motivating additional research on the material and its applicability as a SSE.
As we will show in this work, the ionic conductivity predicted both via accurate, if short, first-principles molecular dynamics as well as extensive classical molecular dynamics is of high value, marking this structure as a  promising candidate SSE.
We calculate its electrochemical stability windows from first principles and find the material to have a wide stability window, comparable to the garnet \ce{Li7La3Zr2O12} (LLZO). 
Cyclic voltammetry and chemical reactivity experiments confirm these findings.
Furthermore, we use impedance spectroscopy to determine the ionic conductivity of sintered \ce{Li7TaO6} pellets, and obtain barriers of $\sim$0.6~eV, lower than reports in the literature, but higher than seen in simulation.
However, we also find evidence of side phases that allow us to rationalize the lower diffusion in experiment with respect to the results of simulations.

Details of the methods employed, both computational and experimental, are given in \mysecref{sec:methods}, followed by the illustration and discussion of the results in \mysecref{sec:results}. 
We summarize our work and present the conclusions in \mysecref{sec:conclusions}.

\section{Methods}
\label{sec:methods}

\subsection{Molecular dynamics simulations}

\begin{figure}[b]
\includegraphics[width=\hsize]{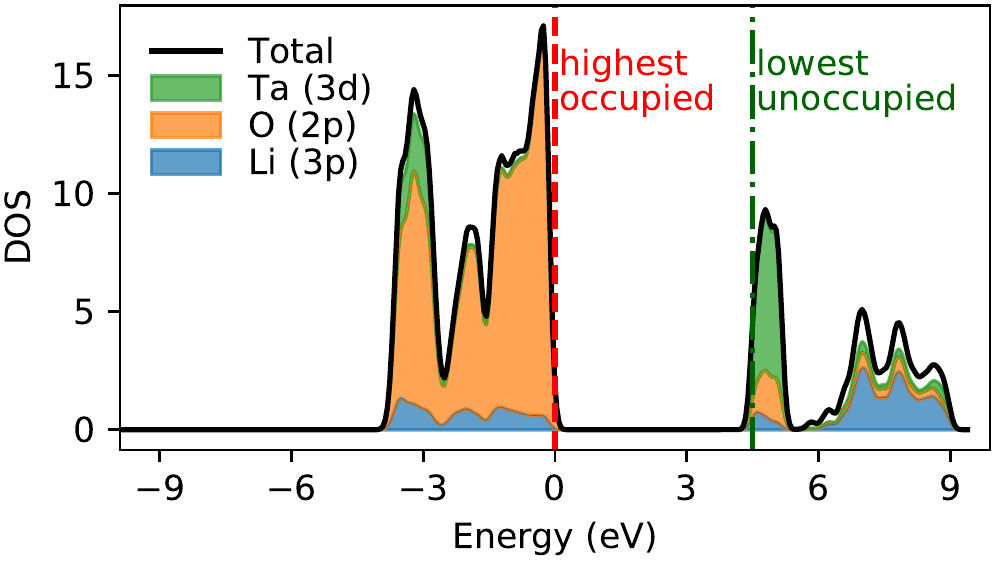}
\caption{
The electronic density of states from DFT+$U_{sc}$ as a black solid line at $\pm 10$~eV from the highest occupied state, marked with a red dashed line at 0~eV (green dash-dotted line for loweest valence state).
We also project the density of states onto the atomic orbitals, shown in green for Ta (3d), orange for O (2p), and blue for Li (3s).
We applied a Gaussian broadening ($\sigma=0.01$~eV) to smoothen the spectrum.
}
\label{fig:tantalate-pdos}
\end{figure}

\paragraph*{First-principles molecular dynamics} (FPMD) are a powerful tool to study the diffusion mechanisms of  solid-state Li-ionic conductors, calculating forces  on the fly and accurately from the ground-state electronic structure at every step during the atomic dynamics.
In this work we performed Born-Oppenheimer FPMD simulations in the framework of density-functional theory (DFT)~\cite{kohn_self-consistent_1965}, using the implementation of the plane-wave pseudopotential method in the PWscf module of the Quantum ESPRESSO distribution~\cite{giannozzi_quantum_2009}.
We used the Perdew-Burke-Enzerhof (PBE)~\cite{perdew_generalized_1996} exchange-correlation functional, and pseudopotentials and cutoffs suggested by the Standard Solid-State Pseudopotential (SSSP) Efficiency library~1.0~\cite{prandini_precision_2018,kucukbenli_projector_2014,dal_corso_pseudopotentials_2014,garrity_pseudopotentials_2014}.

We took the primitive cell of \ce{Li7TaO6} from ICSD~\cite{belsky_new_2002} entry 74950 and
performed a variable-cell relaxation of the primitive cell with a dense $8\times 8 \times 7$ k-point mesh (further details are given \supplsecref{sup-subsec-vc-relax}).
One calculation of the electronic density of states (shown in \myfigref{fig:tantalate-pdos}) using DFT+$U_{sc}$, where the Hubbard $U$ was calculated self-consistently from linear response~\cite{cococcioni_linear_2005,timrov_hubbard_2018} to account for possible strong localization of electrons in the $d$ states of Ta, gave evidence that \ce{Li7TaO6} is an electronic insulator, with a calculated band gap of 4.5~eV at DFT-PBE+$U_{sc}$ level.
For the FPMD simulations we created a $2\times 2\times 2$ supercell of \ce{La7TaO6} (112 atoms) to minimize spurious periodic image interactions and to allow for finite-temperature sampling. 
The resulting supercell will be referred to as \ce{Li56Ta8O48} in the remainder, for clarity.
Due to the presence of a large band gap, the Brillouin zone was sampled at the $\Gamma$ point only, with no electronic smearing, using a threshold for the electronic charge density minimization of $1.12\times 10^{-10}$~Ry.
We performed molecular dynamics simulations using the standard Verlet integrator~\cite{verlet_computer_1967}, with 1.45~fs timesteps, sampling the canonical (NVT) ensemble, i.e. fixing  the number of particles, the volume, and the temperature of the system.
The latter was controlled with a stochastic velocity rescaling thermostat~\cite{bussi_canonical_2007}, implemented by us into PWscf, with a characteristic time for the thermostat set to 0.2~ps, which was found to result in efficient thermalization, but does not affect on the dynamics of the system.
We simulated \ce{Li7TaO6} at  500~K, 600~K, 750~K, and 1000~K, for simulation times of 552~ps, 639~ps, 203~ps, and 73~ps, respectively.
The criterion for stopping the simulations was given by the relative standard error of the diffusion coefficient becoming 10\% or less.
We managed all the dynamics with workflows of the AiiDA materials informatics~\cite{pizzi_aiida:_2016} platform, to ensure automation and fully reproducible results.

\paragraph*{Force-field molecular dynamics:}
 FPMD simulations have accurate interatomic forces but are restricted to short time and length scales.
 In order to support the statistical relevance of our results we developed a classical core-shell potential~\cite{dick_theory_1958} which is suitable to describe  solid-state electrolyte materials~\cite{mottet_doping_2019}.
The inter-atomic potential energy is given by the sum of the electrostatic energy between ions, a Buckingham term describing the short range repulsion, and van-der-Waals interactions between particles $i$ and $j$ at a distance $r_{ij}$:
\begin{equation}
U(r_{ij})=\frac{q_iq_je^2}{r_{ij}}+Ae^{-\frac{r_{ij}}{\rho}}-\frac{C}{r_{ij}^6},
\label{eq:pot-pff}
\end{equation}
where $q$, $A$, $C$, and $\rho$ are species-dependent model parameters. 
In addition to the interactions given in \myeqref{eq:pot-pff}, selected types of atoms can be refined considering them as two particles, referred to as core and  shell, modeling the ionic core and electronic shell respectively.
In these cases the total atomic charge is split between the core and the the shell, allowing for a finite dipole contribution which models polarization effects; we will therefore refer to this classical potential as a polarizable force field (PFF).
The core and the shell interact via harmonic springs, and the mass of the shell must be chosen small enough to ensure adiabatic separation (i.e. no mutual thermalization) between the slow ionic motion and the fast relaxation of the electrons, as is done in the Born-Oppenheimer approximation, and a condition which was always respected in our simulations.

We optimized the parameters of the force-field  following the approach described by Zipoli and Curioni~\cite{fzi} via a simultaneous minimization of the force and energy mismatches on selected configurations, which were selected from short molecular dynamics runs at the PBE level ~\cite{perdew_generalized_1996} including van-der-Waals correction within the Grimme-D2 parametrization ~\cite{grimme_semiempirical_2006} as implemented in the CP2K code~\cite{hutter_cp2k_2014}.
In the present parametrization of \ce{Li7TaO6} only the more polarizable oxygen atoms were treated with a core and a shell;
in \supplsecref{sup-subsec-pff} we report in detail the fitting procedure and its quality.
With the optimized parameters we performed PFF simulations using the LAMMPS code for large scale molecular dynamics simulations~\cite{Plimpton1995},
for a supercell of 90720 atoms.
Access to such long length and time scales permitted to follow a three-step protocol to ensure the reliability of the resulting trajectories.
First, we equilibrated the simulation cell by coupling the entire system (ions and shells) to a barostat, setting the pressure to $p=1$~bar, and coupling the ions to a Berendsen thermostat.
This setup reproduces closely an NPT ensemble in which the light cold shells follow adiabatically the ions, which are coupled to a thermostat.
We performed ten simulations at different target temperatures between 300~K and 900~K.
In a second step we fixed, for each temperature, the equilibrium volume, and the simulations continued in the canonical ensemble (NVT), using a Nos\'e-Hoover integrator with a coupling time of 0.5~ps applied to the ions (no thermostat on the shells).
We note that the temperature of the O shells, about 1~K, presented a negligible drift of $1.5 \times 10^{-2}$~K\,ns$^{-1}$, confirming the adiabatic decoupling between the motion of ions and shells. 
Finally, the thermostat was removed to sample at constant volume and energy in the microcanonical (NVE) ensemble.
This way we ruled out the possibility that the thermostat or barostat influenced the dynamics, and we used the NVE trajectories to calculate diffusion coefficients. The length of the NVE simulations depended on the target temperature, and both are reported in the \supplsecref{subsec-addition-pff-results}.

\paragraph*{Analysis of trajectories:}
The mean-square displacements (MSD) of the diffusing species is the microscopic property that links to the (macroscopic) tracer diffusion coefficient~\cite{einstein_uber_1905}. We calculated the MSD of species  $S$, with $N_S$ atoms in the simulation cell, from the molecular dynamics trajectory:
\begin{equation}
    \mathrm{MSD}^S(\tau) = \frac{1}{N_S} \sum_i^S   \big\langle | \bm r_i(\tau+t)-\bm r_i(t))|^2\big\rangle_t,
    \label{eq:msd}
\end{equation}
where $\bm r_i(t)$ is the position of the $i$-atom 
at time $t$.
Averaging over all trajectory timesteps is denoted as a time average by the angular brackets $\langle\cdots\rangle_t$; assuming ergodicity, this is equivalent to an ensemble average under the relevant thermodynamic conditions.
The tracer diffusion coefficient of species $S$,  $D_{tr}^S$, was calculated from the long time limit of the MSD:
\begin{equation}
    D_{tr}^S = \lim_{\tau \rightarrow \infty} \frac{1}{6\tau} \mathrm{MSD}^S(\tau).
    \label{eq:diff-isotropic}
\end{equation}
We performed a least-squares regression to find the line of best fit to the MSD in the diffusive regime, while block analysis~\cite{allen_computer_1987} was used to estimate the error of the tracer diffusion coefficient.
Via the Nernst-Einstein equation, and assuming a Haven ratio of one, we estimated the ionic conductivity $\sigma$:
\begin{equation}
\sigma=\frac{e^2C}{k_BT}D_{tr}^\mathrm{Li},
\label{eq:nernst-einstein}
\end{equation}
where $C$ is the average Li-ion density in the system.
We also estimated the tracer diffusion matrix  $\bm D^{S}$ of species $S$ as follows:
\begin{equation}
 D_{ab}^S = \lim_{\tau \rightarrow \infty} \frac{1}{2\tau}  \frac{1}{N_S} \mathlarger{\sum}_i^S\Big\langle\big(r^a_i(\tau+t)-r^a_i(t)\big) \cdot 
 \big(r^b_i(\tau+t)-r^b_i(t)\big) \Big\rangle_t,
\label{eq:diff-anisotropic}
\end{equation}
where $r_i^a(t)$ is the position of the $i$-th atom along the Cartesian coordinate $a$ at time $t$.
The tracer diffusion coefficient defined in \myeqref{eq:diff-isotropic} can also be calculated from the diffusion coefficient matrix of \myeqref{eq:diff-anisotropic} as: $D_{tr}^S=1/3 \cdot \mathrm{Tr} (\bm D^{S})$.
$\bm D^{S}$ contains additional information on the anisotropy of the diffusion.
The normalized variance of the eigenvalues of the diffusion matrix is referred to as the fractional anisotropy (FA)~\cite{fractional_aniso}.

The Li-ion probability density $n_\text{Li}({\bm r})$ was calculated from trajectories via an average over all the frames of the trajectory:
\begin{equation}
 n_\text{Li}(\bm r) =\left\langle \sum_i^{\text{Li}} \delta(\bm r-\bm r_i(t)) \right\rangle_t,
 \label{eq:density}
\end{equation}
where the Li-ions are indexed by $i$.
Due to finite statistics, we replaced the delta function in \myeqref{eq:density} by a Gaussian with a standard deviation of 0.3~\r{A}.

We also calculated the radial distribution function (RDF) 
$g(r)_{S-S'}$ of species $S$ with species $S'$ as:
\begin{equation}
 g_{S-S'}(r) = \frac{\rho(r)}{f(r)} = \frac{1}{f(r)} \frac{1}{N_S} \sum_i^S
\sum_j^{S'}\Big\langle \delta \left(r - \left| \bm r_i(t) - \bm r_j(t) \right| \right) \Big\rangle_t,
\label{eq:rdf}
\end{equation}
where $f(r)$ is the ideal-gas average number density at the same mean density. 
We obtained the coordination number at $r$  by integrating the average number density $\rho(r')$ from 0 to $r$.

In addition, we analyzed the trajectories using the \textsc{sitator} package, allowing for an unsupervised analysis~\cite{kahle_unsupervised_2019} of the resulting sites and diffusive pathways.
The analysis returns states and transitions of the diffusion pathway by projecting the Li-ion coordinates into a finite-dimensional vector space describing the position relative to the host lattice, in this case the tantalum and oxygen sublattice of \ce{Li7TaO6}.
A subsequent clustering permits to identify crystallographic sites.
The parameters used in this work are given in \supplsecref{sup-subsec-sitator}.

\subsection{Electrochemical Stability}
\label{subsubsec-ec-stab-comp-method}
The electrochemical stability of a solid-state electrolyte (SSE) material describes its robustness against either reduction (at the low potential electrode interface) or oxidation (at the high potential electrode interface) and is quantified in terms of the stability potential window, which is limited by the SSE reduction potential $\Phi_{\text{red}}$ and oxidation potential $\Phi_{\text{ox}}$ from below and above, respectively.
The most important processes that limit SSE stability are Li-exchange reactions at the interfaces between the electro-active materials of the electrodes and the SSE. 
We distinguish two different types of reactions~\cite{binninger_arxiv_2019}: those that result in major SSE phase decomposition~\cite{mo_first_2012} and those that alter the SSE stoichiometry but leave the SSE structure intact~\cite{nakayama_first-principles_2012}.
The first type defines the potential window of phase stability and the second type the potential window of stable stoichiometry~\cite{binninger_arxiv_2019}.
Each Li-exchange reaction defines an equilibrium potential, given by the corresponding Gibbs free energy of reaction, where Li-ion transfer from electro-active material to SSE represents a reduction of the SSE, and where Li-ion transfer from the SSE to the electro-active material represents SSE oxidation.
The relevant SSE reduction potential $\Phi_{\text{red}}$ is given by the maximum of all reduction reaction potentials, and the relevant SSE oxidation potential $\Phi_{\text{ox}}$ is given by the minimum of all oxidation reaction potentials.

We assessed the electrochemical stability of \ce{Li7TaO6} with two different computational methods, the details of which have been presented in previous work~\cite{binninger_arxiv_2019}.
Gibbs free energies of reaction were approximated by the pure reaction energies, which were computed from density-functional theory (DFT) using the Quantum ESPRESSO~\cite{giannozzi_quantum_2009} software package.
Also here we used the PBE functional~\cite{perdew_generalized_1996} and the same pseudopotentials from the SSSP Efficiency 1.0 library~\cite{prandini_precision_2018}, but the plane-wave basis energy cutoff was set to 100~Ry for the wavefunctions and 800~Ry for the electronic charge density.
Due to the large supercell sizes of $\sim$1500~\r{A}$^3$, k-point sampling could be restricted to the $\Gamma$-point.

\subsection{Experiments}
\label{subsec-experimental}
\paragraph*{Synthesis:} We synthesized \ce{Li7TaO6} with a solid-state reaction.
In a first synthesis process 
stoichiometric amounts of \ce{Ta2O5} (Alfa-Aesar 99.993\%) and \ce{Li2O} (Alfa-Aesar 99.5\%) were ground together.
In a second  process, we used an excess of \ce{Li2O}, corresponding to an overall stoichiometry Li:Ta\,=\,8:1, to compensate for possible Li loss during synthesis~\cite{nomura_ionic_1984}.
The product of this latter synthesis approach is denoted `xs-\ce{Li7TaO6}' in the following. 
After grinding, we pressed the mixture into a cylindrical pellet, 13~mm in diameter and about 5~mm thick, at 5~tons load.
Afterwards, the pressed pellet was heated at 700~$^{\circ}$C for 48~hours on a commercially available MgO single crystal wafer in an alumina crucible under a dry argon atmosphere.
For \ce{Li_{7+x}Ta_{0.98}Zr_{0.02}O6} and \ce{Li_{7+x}Ta_{0.98}Mo_{0.02}O6}, stoichiometric amounts of \ce{ZrO2} (Sigma-Aldrich 99\%) and \ce{MoO3} (Sigma-Aldrich 99.5\%) powders were added into the \ce{Ta2O5} and \ce{Li2O} mixtures, respectively.
 We also prepared pellets of high relative density of \ce{Li7TaO6} using a commercial spark plasma sintering (SPS) system at 950$^{\circ}$C with a holding time of 5~minutes in a 10~mm-diameter graphite mold, under an axial compressive stress of 100~MPa,  and under dynamic primary vacuum.

\paragraph*{XRD and ICP-OES:}
After cooling, we crushed the pellets and identified the phases present via powder X-ray diffraction (XRD, Bruker D8 system) with Cu~K-$\alpha$ polychromatic radiation ($\lambda=0.15418$~nm) in a Bragg-Brentano geometry for 1~h from 10$^\circ$ to 90$^\circ$ with a step size of 0.0334$^\circ$.
For the refinement, we collected X-ray powder diffraction data of \ce{Li7TaO6} at room temperature for 12~h from 11$^\circ$ to 70$^\circ$ with a step size of 0.0334$^\circ$, using an X-Pert Panalytical $\mathrm{\theta}$-2$\mathrm{\theta}$ Bragg geometry diffractometer (monochromatic copper source $\lambda = 1.54056$~\r{A}). The Rietveld refinement was performed using the FULLPROF software~\cite{rodriguez_iucr_2001}.
We determined the sample stoichiometry after synthesis using Inductively Coupled Plasma Optical Emission Spectrometry (ICP-OES). 

\paragraph*{X-ray photoelectron spectroscopy:} We performed XPS measurements using a VG~ESCALAB~220iXL spectrometer (Thermo Fischer Scientific) with a monochromatized Al K-$\alpha$ source (1486.6~eV; beam size $\sim$500~$\mathrm{\mu m^2}$; 150~W) and the following conditions/parameters:
$2 \times 10^{-9}$~mbar analysis chamber pressure; 30~eV pass energy (50~eV for surveys) in 50~meV steps with 50~ms dwell time~\cite{ferraresi_sno2_2018}.
To avoid/minimize exposure to air and moisture, all XPS samples were prepared in an Ar-filled glovebox and loaded into the spectrometer using an air-tight transfer chamber.
We sputtered the samples with Ar inside the XPS chamber, where a beam energy of 3~kV was applied using an IQE~11/35 ion source (SPECS) under $10^{-5}$~mbar of Ar pressure.
Regarding peak deconvolutions, Shirley-type background subtraction was applied followed by the sum of 70\%~Gaussian -- 30\%~Lorentzian line shapes.
All sample spectra were calibrated to 286~eV, which corresponds to the C1s hydrocarbon component. 

\paragraph*{Electrochemical impedance spectroscopy:}
We prepared pellets for ionic conductivity measurements by pressing the powder to 13~mm diameter and about 2~mm thickness at 5~tons load, followed by sintering at 1100~$^{\circ}$C for 2~hours in \ce{O2}.
The relative density of the pellet was calculated from the measured mass, thickness, and area relative to the theoretical \ce{Li7TaO6} density of 4.25~g\,cm$^{-3}$.
We polished both surfaces of the pellets with silicon carbide (P500) paper and sputtered these with about 100~nm of Ti, followed by another sputtering of 100~nm of Pt.
We performed electrochemical impedance spectroscopy (EIS) measurements  using a Solatron 1260 gain-phase analyzer between 100~mHz and 1~MHz in Ar while cooling the sample from 200~$^{\circ}$C to 40~$^{\circ}$C at intervals of 20~$^{\circ}$C.
At each temperature, samples were stabilized for 1~hour before acquiring the impedance spectra. 

\paragraph*{Electronic conductivity.}
We measured the room-temperature electronic conductivity of \ce{Li7TaO6} using a previously reported technique~\cite{han_high_2019}, where
both surfaces of the polished \ce{Li7TaO6} pellet 
are sputtered with 100~nm Cu before assembly in a Swagelok cell.
A 100~mV DC voltage was applied to the Cu/\ce{Li7TaO6}/Cu cell until a steady state was reached, wt which point the DC current was attributed to pure electronic leakage.  

\paragraph*{Electrochemical measurements.}
For all electrochemical tests, we assembled the \ce{Li7TaO6} pellets in Swagelok cells and cycled them with a VMP-3 multi-channel potentiostat/galvanostat (Bio-logic) in a temperature-controlled chamber (ESPEC SU-221) at 80~$^{\circ}$C.
First, the \ce{Li7TaO6} pellets 
were dry-polished with multiple grit papers (80, 320, 500, then 1200) until a thickness of approximately $500-600~\mathrm{\mu}$m was achieved. 
To improve the contact and wettability with Li foil, each side of the polished pellet was sputtered with 100~nm Au. We conducted cyclic voltammetry (CV) measurements in a semi-blocking cell configuration (Au-\ce{Li7TaO6}-Au/Li), with the working electrode at the Au side and the counter+reference electrode at the Au/Li side, at a relatively slow scan rate of 0.1~mV\,s$^{-1}$ over a potential range between -0.5~V to 6~V  vs. Li$^+$/Li to elucidate the electrochemical stability windows of \ce{Li7TaO6}. 
In addition to CV measurements, we investigated the direct reactivity of \ce{Li7TaO6} with metallic Li.
A polished \ce{Li7TaO6} pellet was assembled in contact with Li foil inside a Swagelok cell and then kept at 80~$^{\circ}$C for 5~hours.
After exposure to metallic Li, we employed XPS to probe the \ce{Li7TaO6} surface.


\section{Results and discussion}
\label{sec:results}
\subsection{Li-ion diffusion in MD simulations}
\label{sec-res-diffusion}

\begin{figure}[t]
    \centering
    \includegraphics[width=\hsize]{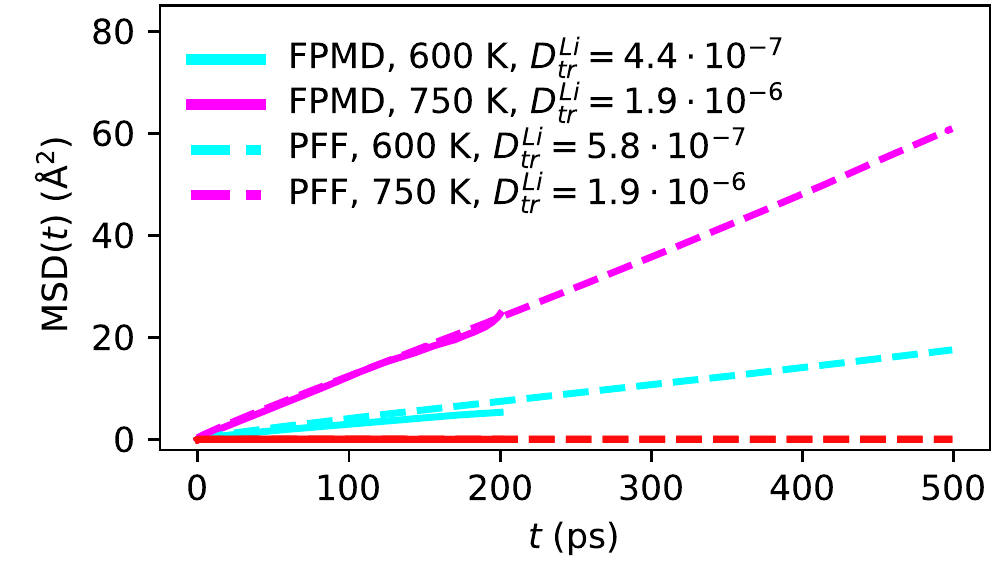}
        \caption{
        The MSD(t) of Li and O, calculated in the small \ce{Li56Ta8O48} supercell, at 600~K (cyan) and 750~K (purple) from the PFF and FPMD simulations is given by dashed and solid lines, respectively.
Oxygen, in red, shows no diffusion in any simulation, evidenced by a flat MSD.
The diffusion of Li ions, estimated from the slope of the MSD, is given in the legend (in cm$^2$\,s$^{-1}$).}
    \label{fig:msd4}
\end{figure}

\begin{figure}[b]
    \centering
    \includegraphics[width=\hsize]{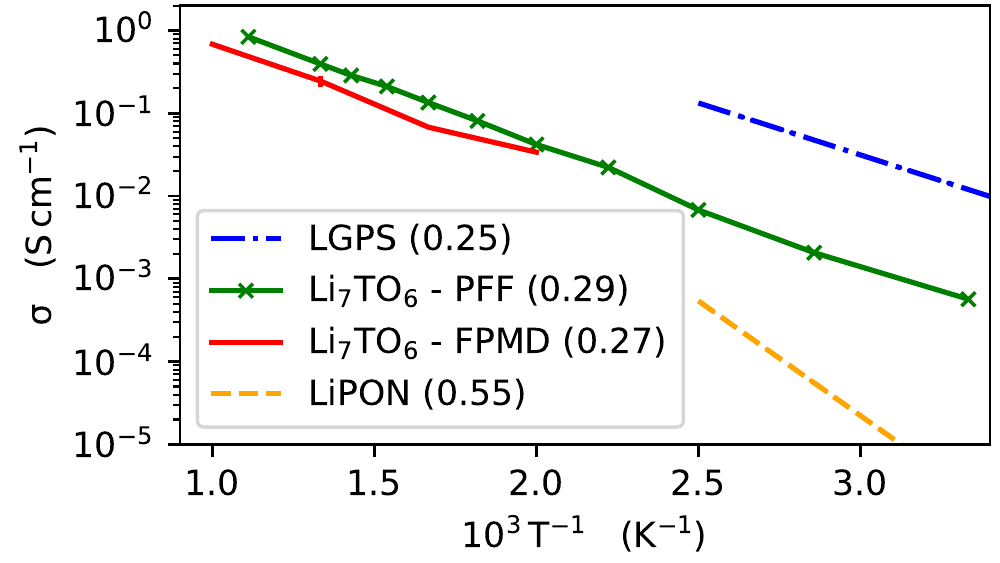}
    \caption{Arrhenius behavior of the ionic conductivities of \ce{Li7TaO6} simulated with FPMD (red solid line) and PFF (green solid line).
    We compare to the experimental ionic conductivities of LGPS~\cite{kamaya_lithium_2011} (blue dashed-dotted line) and LiPON~\cite{yu_stable_1997} (orange dashed line).
    The activation energies are given in the legend within brackets (in eV).
}
    \label{fig:ionic-tantalate-comp}
\end{figure}

\begin{figure}[t]
    \centering
    \includegraphics[width=\hsize]{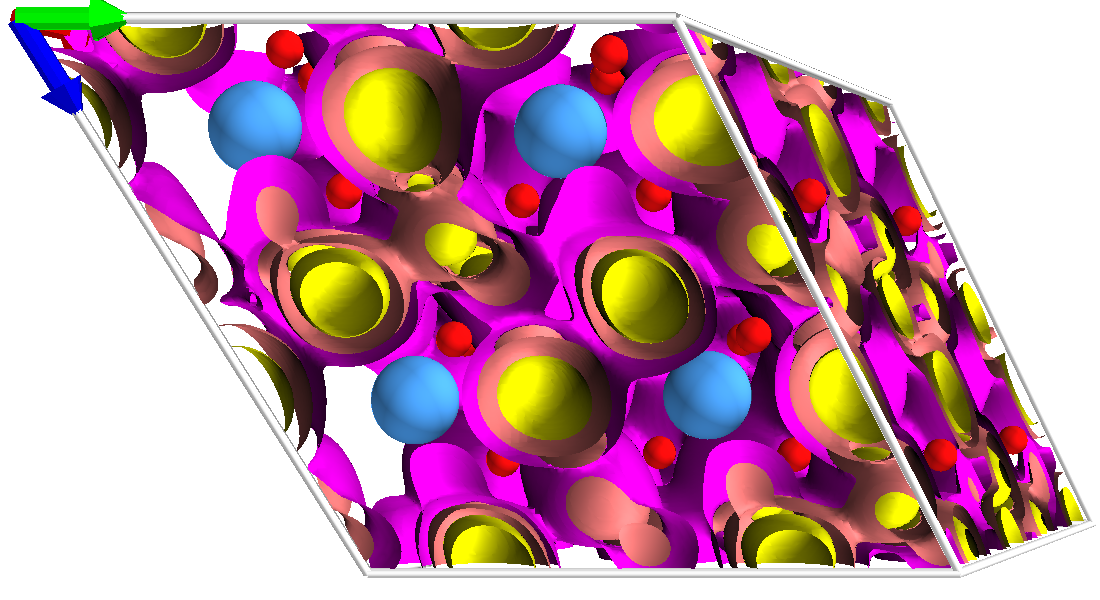}
    \caption{
The Li-ion density of \ce{Li7TaO6} at 600~K.
We show the isosurfaces at 0.001, 0.01, and 0.1~\r{A}$^{-3}$ in violet, orange, and yellow, respectively.
The average positions of the oxygens are shown as red spheres, and of tantalum as blue spheres. 
We chose a different orientation compared to Figs.~\ref{fig:tantalate} and \ref{fig:site-analysis}, to showcase more clearly the diffusion in the plane of Ta ions and out of plane.
}
    \label{fig:density-lto-600}
\end{figure}

\begin{figure}[t]
    \centering
    \includegraphics[width=\hsize]{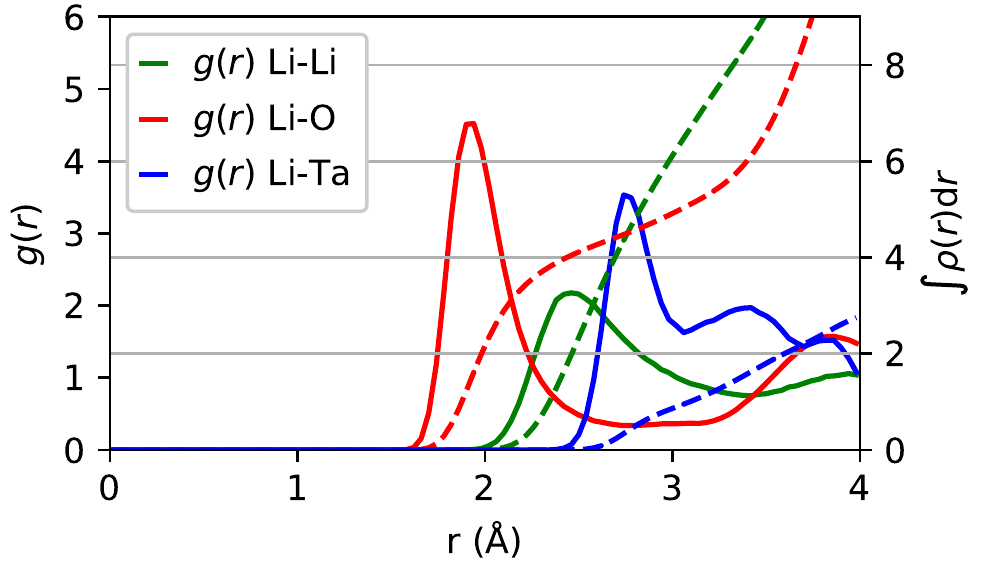}\\
    \includegraphics[width=\hsize]{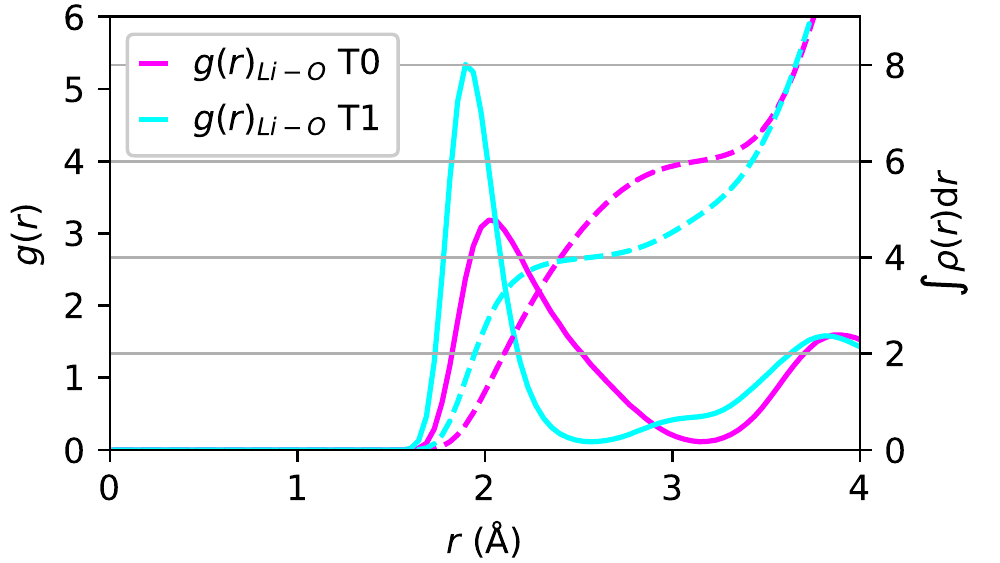}
    \caption{
(Top) The Li-Li, Li-O, and Li-Ta RDF from FPMD trajectories at 600\,K are shown as green, red, and blue solid lines, respectively.
We show the integral of average number density as dashed lines, using the same color encoding.
(Bottom) The Li-O RDF for site type 0 in violet and site type 1 in cyan.
}
    \label{fig:rdf-lto-600}
\end{figure}

\begin{figure}[t]
    \centering
    \includegraphics[width=\hsize]{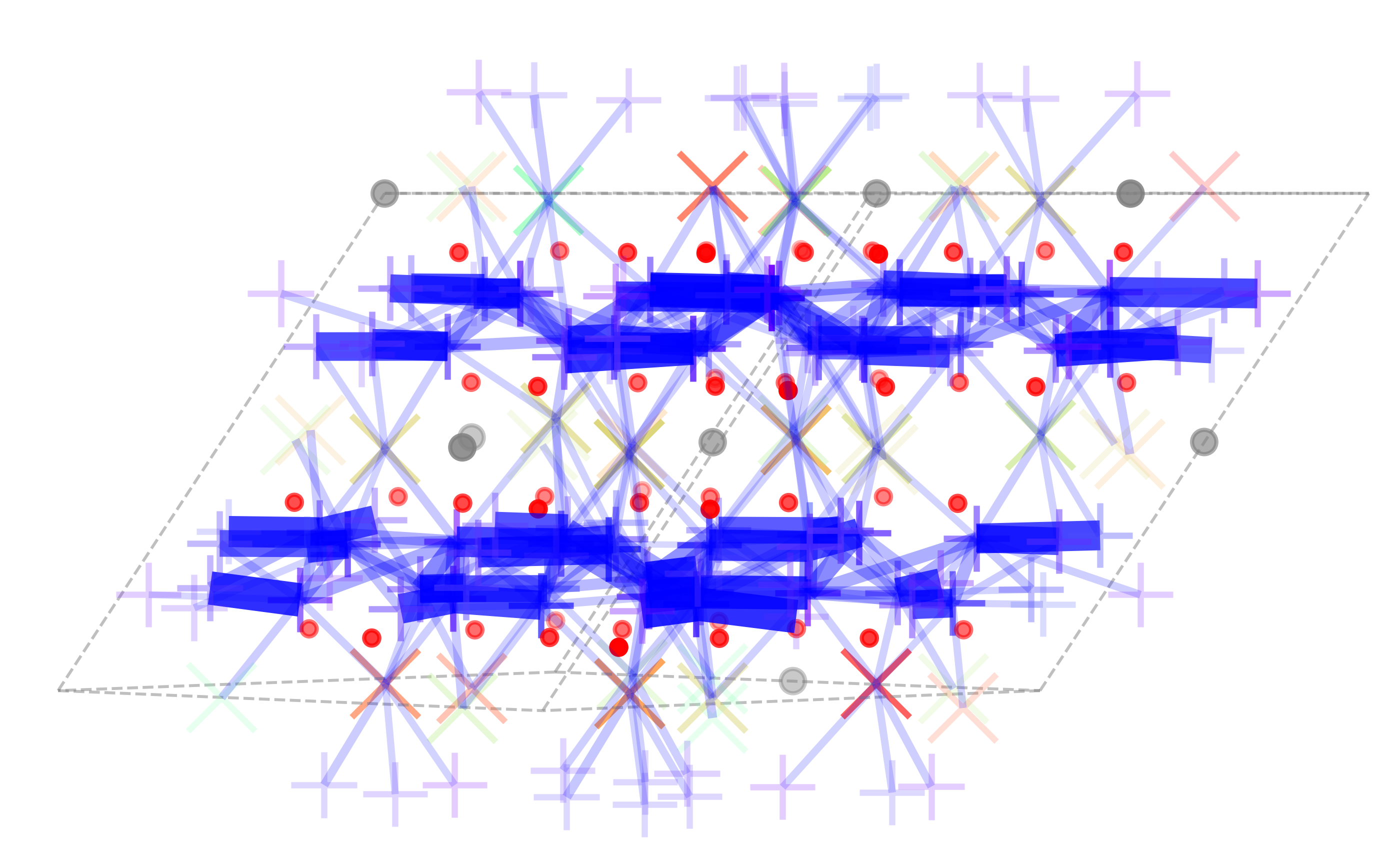}
    \caption{
    We show the results of the site analysis for a $2\times 2\times 2$ cell of \ce{Li7TaO6}.
    Equilibrium positions of oxygen and tantalum are shown as red and grey spheres, respectively.
	The octahedral sites (type 0) for Li ions are given by crosses
	and the tetrahedral sites (type 1) by plusses; the lines connecting the sites are drawn if these sites have exchanged Li-ions, with the thickness indicating the observed flux.
    }
    \label{fig:site-analysis}
\end{figure}

First, the compatibility between FPMD and PFF simulations with respect to the Li-ionic diffusion is assessed for the same  \ce{Li56Ta8O48} supercell, each relaxed at 0~K prior to the simulation.
In \myfigref{fig:msd4} we plot the MSD of Li and O, calculated using \myeqref{eq:msd} from the FPMD and PFF simulations at 600~K and 750~K. 
%
A diffusive regime can be clearly discerned, evidence that the FPMD simulations performed are sufficiently long to get converged results.
The MSD calculated from the FPMD (solid lines) and PFF (dashed lines) simulations give compatible slopes.
The MSD of the oxygen atoms gives evidence for the stability of the rigid framework formed by the non-diffusive atoms and for the absence of lattice drift.
Possible size effects, discussed in greater detail in section \supplsecref{subsec-sup-bulk-diff}, are not expected to change significantly the conductivity in the simulations.
The excellent agreement between FPMD and PFF allows us to use the PFF to calculate the ionic conductivity also at lower temperatures, for a more rigorous prediction at ambient temperature, which requires longer simulation times.
We use the PFF in NPT simulations to equilbrate the lattice parameters at the target temperature, and we control for size effects by using a larger supercell.
In \myfigref{fig:ionic-tantalate-comp} we report the computed ionic conductivities from the PFF and FPMD simulations together with
the conductivities of LGPS~\cite{kamaya_lithium_2011} and LiPON~\cite{yu_stable_1997} for comparison.
We estimate the activation energy for \ce{Li7TaO6} at
0.27~eV from the FPMD simulations and 0.29~eV from the PFF simulations, in good agreement with each other.
The activation energies for \ce{Li7TaO6} are slightly higher than the experimental activation barrier for LGPS, which ranges from 0.22 to 0.25~eV~\cite{kuhn_tetragonal_2013,kamaya_lithium_2011} and much lower than the activation barrier of 0.55~eV found in LIPON~\cite{yu_stable_1997}.
Overall the results of our simulations indicate that \ce{Li7TaO6} could be a very promising Li-ionic conductor, suitable for application as solid-state electrolyte.

The Li-ion probability density $n(\bm r)$ from the FPMD simulation at 600~K (our longest FPMD trajectory), is reported in \myfigref{fig:density-lto-600},
where we show three different isosurfaces.
The lower isovalue of 0.001~\r{A}$^{-3}$ allows to discern a connected network of Li-ion diffusion pathways, additional evidence that the material is a good ionic conductor.
The high-density isovalue displays disconnected regions of high probability density, showing that crystallographic sites can be indeed clearly identified.

From the FPMD trajectories we calculated the RDF of Li with all other species between 0~and 4~\r{A}, using \myeqref{eq:rdf}.
The resulting RDFs, shown in \myfigref{fig:rdf-lto-600}, reveal that Li ions are, as expected, closest coordinated by oxygen, but the coordination number cannot be rigorously extracted due to the fact that this type of analysis is not able to distinguish between different types of sites for Li in different coordination.
To understand further the geometry of the Li sites we see in simulation, we analyzed the Li-ion dynamics with the \textsc{sitator} package~\cite{kahle_unsupervised_2019}, to characterize the distinct sites ions visit during the simulation.
By computing chemical and geometrical fingerprints for each site, we classify sites by their type.
We find two sites that we label type 0 and six sites that we label type 1, per formula unit of \ce{Li7TaO6}, for both the FPMD trajectory and the PFF trajectory at 600~K.
The clustering of the descriptor vectors for each site (see \supplfigref{fig:soaps}) results in distinct clusters, evidence of a clear difference in the geometric and chemical environment of the sites. 
Calculating the $g(r)_{Li-O}$ for each site type separately we observe a different coordination number, which is estimated from the plateau of the Li-O RDF shown in the bottom panel of \myfigref{fig:rdf-lto-600}.
Site type 0 is coordinated by six oxygen atoms and is an octahedral site, compared to four oxygen atoms for the tetrahedral site type 1, in excellent agreement with published results~\cite{wehrum_zur_1994, delmas_conducteurs_1979} on the Li-ion sites in this structure.
\myfigref{fig:site-analysis} shows the connectivity of the sites for a PFF trajectory at 600~K.
From this analysis we see one connected component, i.e. starting from every site, a Li-ion can reach every other site in the system via a sequence of jumps,
but we observe that the material has a more strongly connected plane of diffusion between the octahedral sites of type 0.
There is however also significant diffusion perpendicular to the plane, where the tetrahedral sites are located.
This behavior is reflected in the value of the FA, which is stable with respect to the temperature at a value of $\sim 0.3$, signature of an anisotropic type of diffusion. Ideal FA values are 0.0, 0.7, and 1.0 for isotropic, two-dimensional, and uni-dimensional diffusion.
Analyzing further the diffusion matrix, we find two equal eigenvalues, with the respective diffusion direction lying in plane, whereas the perpendicular diffusion is a factor of $\sim$2 lower.
We therefore classify the material as a three-dimensional conductor, but with a preference for in-plane diffusion.

\subsection{Structure and ionic conductivity from experiment}

\begin{figure}[t]
    \centering
    \includegraphics[width=\hsize]{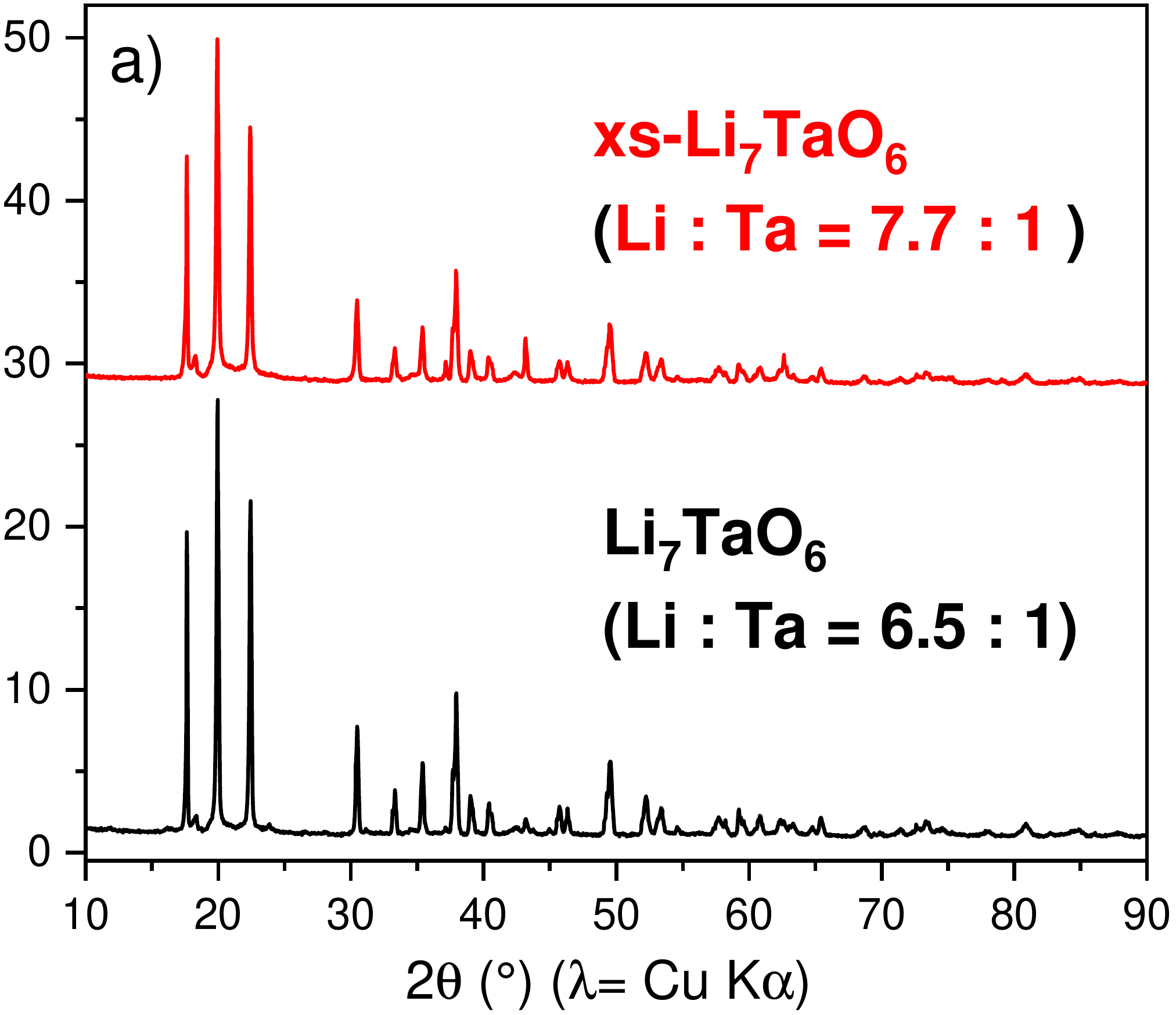}
    \hspace{0.4cm}
    \includegraphics[width=\hsize]{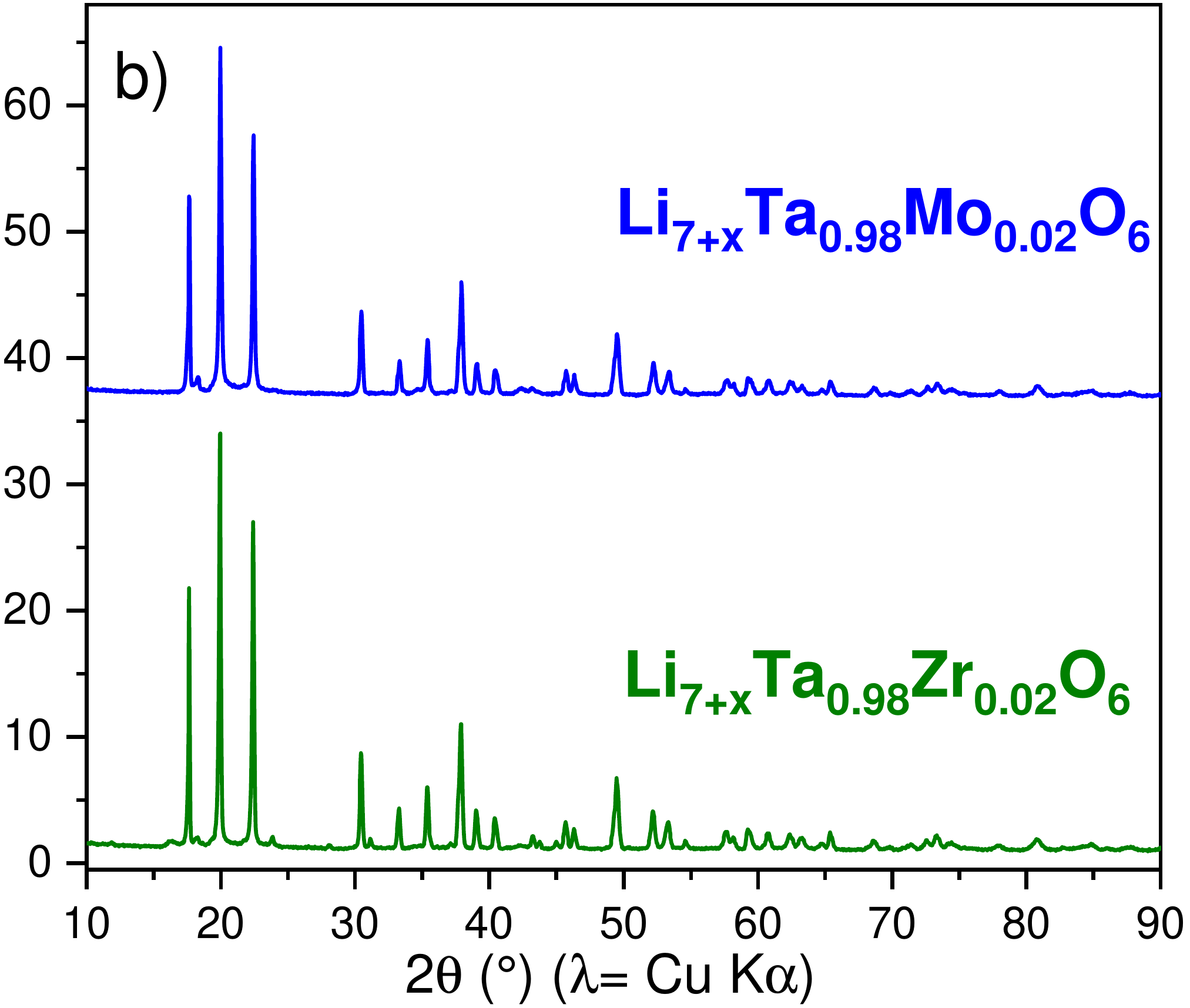}
    \caption{XRD patterns for (a) \ce{Li7TaO6} and xs-\ce{Li7TaO6}; (b) \ce{Li_{7+x}Ta_{0.98}Zr_{0.02}O6} and \ce{Li_{7+x}Ta_{0.98}Mo_{0.02}O6}. The overall stoichiometric compositions shown in parentheses were calculated based on ICP-OES results (2\% relative error).}
    \label{fig:xrd}
\end{figure}

\begin{figure}[b!]
    \centering
    \includegraphics[width=\hsize]{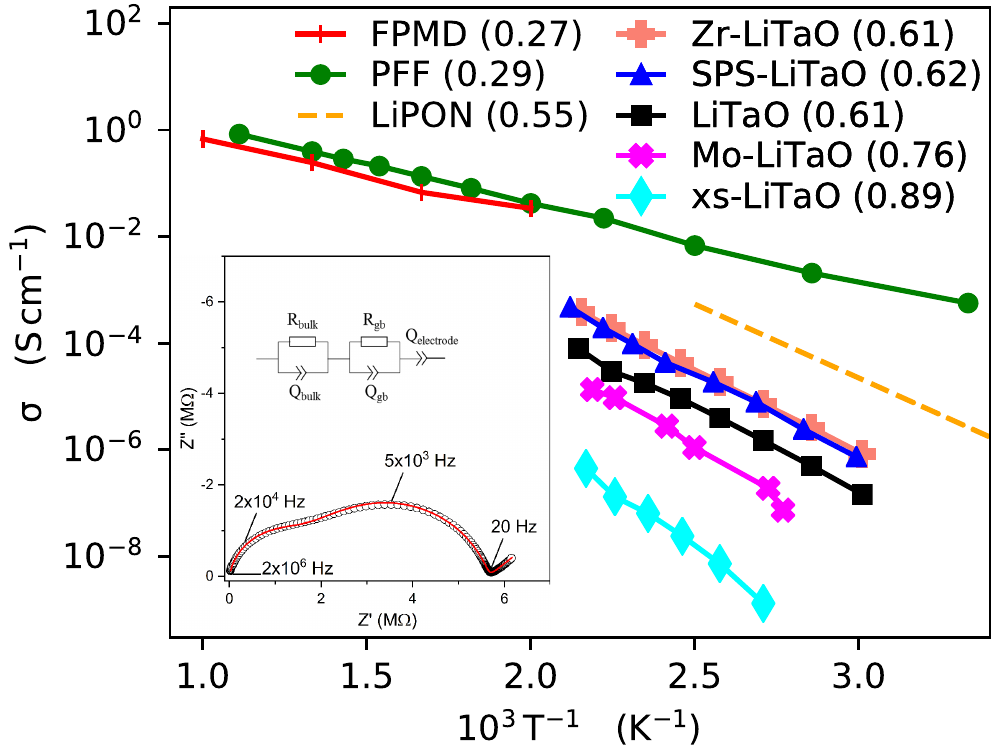}
    \caption{
    Arrhenius plot of experimentally determined ionic conductivities (bulk + grain boundary contributions) of
    xs-\ce{Li7TaO6} (cyan diamonds),
    \ce{Li_{7+x}Ta_{0.98}Mo_{0.02}O6} (purple crosses),
    \ce{Li7TaO6} (black squares),  
    \ce{Li7TaO6} prepared via SPS (blue triangles), and
    \ce{Li_{7+x}Ta_{0.98}Zr_{0.02}O6} (red crosses).
    We compare these to the ionic conductivities from the PFF (green circles), FPMD (red lines), and LiPON experiments~\cite{yu_stable_1997} (orange dashed line).
    The barriers to diffusion are given in the legend within brackets in eV.
	The Nyquist plot of the electrochemical impedance measured in a polished \ce{Li7TaO6} pellet with contributions from both grain boundary and bulk resistance is shown in the inset.    
    }
    \label{fig:ion_cond}
\end{figure}

Historically, the structure of \ce{Li7MO6} (M = Sb, Ti, Nb, Ta) was described by a rhombohedral layer-type structure, formed by hexagonally close-packed oxygen octahedra~\cite{scholder_uber_1958,delmas_conducteurs_1979,hauck_notizen:_2014}.
In the case of \ce{Li7TaO6}, Wehrum and Hoppe~\cite{wehrum_zur_1994} confirmed the P3 space group ($a=5.358$~\r{A}; $c=15.073$~\r{A}).
More recently, using high-resolution synchrotron diffraction data, M\"uhle \textit{et al.}~\cite{muhle_new_2004} observed additional splitting in the reflections, leading to a triclinic unit cell.
We obtain  the best agreement factor for \ce{Li7TaO6} with the P-1 space group, and therefore use this space group  for the  starting parameters of the fit, with atomic positions taken from M\"uhle \textit{et al.}~\cite{muhle_new_2004}.
Atomic positions and the thermal agitation 
factor of oxygen have been fixed voluntarily, 
in agreement with synchrotron data analyzed by M\"uhle \textit{et al.}
A good agreement factor for the refinement (RBragg = 12.5\%, Rwp = 13.3\%; Chi$^2 = 5.28$) confirms that our data are consistent with a triclinic unit cell. 
Table~\ref{tab:cell_parameters} sums up the crystallographic cell parameters  we obtained from the refinements, and we report the atomic positions in \supplref[Table~]{supp-subsec-xrd-res}.

\begin{table}[b]
    \caption{\ Cell parameters from the present XRD results, compared with the results by M\"uhle \textit{et al.}~\cite{muhle_new_2004} as reference} 
    \label{tab:cell_parameters}
    \centering
	\begin{tabular*}{\hsize}{@{\extracolsep{\fill}}lll}
         \hline
         Parameter & Present results & Reference \\ \hline
         a & 5.41 & 5.38486 \\ 
         b & 5.94 & 5.92014 \\
         c &  5.40 & 5.38551 \\
         $\alpha$ & 117.13 & 117.0108 \\
         $\beta$  & 119.82 & 119.6132 \\
         $\gamma$ & 63.06 & 63.2492 \\
         \hline
    \end{tabular*}
    
\end{table}

\myfigref{fig:xrd}a shows the X-ray diffraction (XRD) patterns of \ce{Li7TaO6} and xs-\ce{Li7TaO6}, as well as the Li:Ta stoichiometries of the sample pellets calculated based on  results from Inductively Coupled Plasma Optical Emission Spectrometry (ICP-OES).
Both XRD patterns reveal the same crystalline \ce{Li7TaO6} phase (triclinic metric, space group P-1).
The XRD spectra in \myfigref{fig:xrd}b show that the partial substitution of 2\% Ta by Mo or Zr results in a similar structure as \ce{Li7TaO6}.
However, all samples also produced minor impurities that could not be attributed to the \ce{Li7TaO6} structure, e.g. for $2\theta$ around 18-19$^{\circ}$ and between 41-45$^{\circ}$.
ICP-OES results (with approximately 2\% relative error) reveal a 10\% excess of Li present in the xs-\ce{Li7TaO6} sample. 
We assume that this excess Li is a accommodated by Li-rich side phases, because the \ce{Li7TaO6} phase itself should not become super-stoichiometric in Li due to its negative reduction potential vs. metallic Li, as we will show in the results for the electrochemical stability later on. 
Vice versa, ICP-OES results indicate a Li deficit in the \ce{Li7TaO6} sample.
Due to the large oxidation potential of stoichiometric \ce{Li7TaO6} (cf. electrochemical stability results in \mysecref{subsec-electrochemical-results}) it is unlikely that the \ce{Li7TaO6} phase itself is Li deficient.
Thus, our results suggest that the stoichiometric \ce{Li7TaO6} synthesis suffered from Li-loss resulting in Li-poor side phases. 
A further refinement of the Li excess used in the synthesis appears necessary to obtain \ce{Li7TaO6} at sufficient purity. 

The calculated relative densities of the pellets prepared by the conventional process, i.e. pressing then sintering, and by the SPS method, are shown in Table~\ref{tab:dens}.
A high relative density with a compact microstructure has been reported as being essential for achieving high ionic conductivity in SSE pellets~\cite{chen_influence_2002, wolfenstine_hot-pressed_2010}.
The xs-\ce{Li7TaO6} pellet prepared with the conventional process exhibits lower relative density than the \ce{Li7TaO6} pellet prepared with the same pressing-sintering process, which could be an indication of a stronger presence of side phases in the former sample, due to the \ce{Li2O} excess during synthesis.
The SPS method allows to obtain a very dense pellet of \ce{Li7TaO6} with relative density of 100\%. 
However, the SPS method is not an economical and practical process of pellet preparation.
As an alternative, substituting 2\% Zr for Ta produces a high relative density pellet of \ce{Li_{7+x}Ta_{0.98}Zr_{0.02}O6} by the conventional process.
Conversely, doping with Mo has an opposite effect on relative density.
\begin{table}[b]
\caption{\ Relative densities of pellets prepared with different processes 
}
\label{tab:dens}
\centering
\begin{tabular*}{\hsize}{@{\extracolsep{\fill}}lll}
 \hline
 Compound & Preparation process & Density \\ 
 \hline
 xs-\ce{Li7TaO6} & Press \& sinter & 71\% \\ 
 \ce{Li7TaO6} & Press \& sinter & 84\% \\
 \ce{Li7TaO6} & SPS & 100\% \\
 \ce{Li_{7+x}Ta_{0.98}Zr_{0.02}O6} & Press \& sinter & 93\% \\
 \ce{Li_{7+x}Ta_{0.98}Mo_{0.02}O6} & Press \& sinter & 78\% \\
 \hline
\end{tabular*}
\end{table}

\begin{figure*}[t]
    \centering
    \includegraphics[width=\hsize]{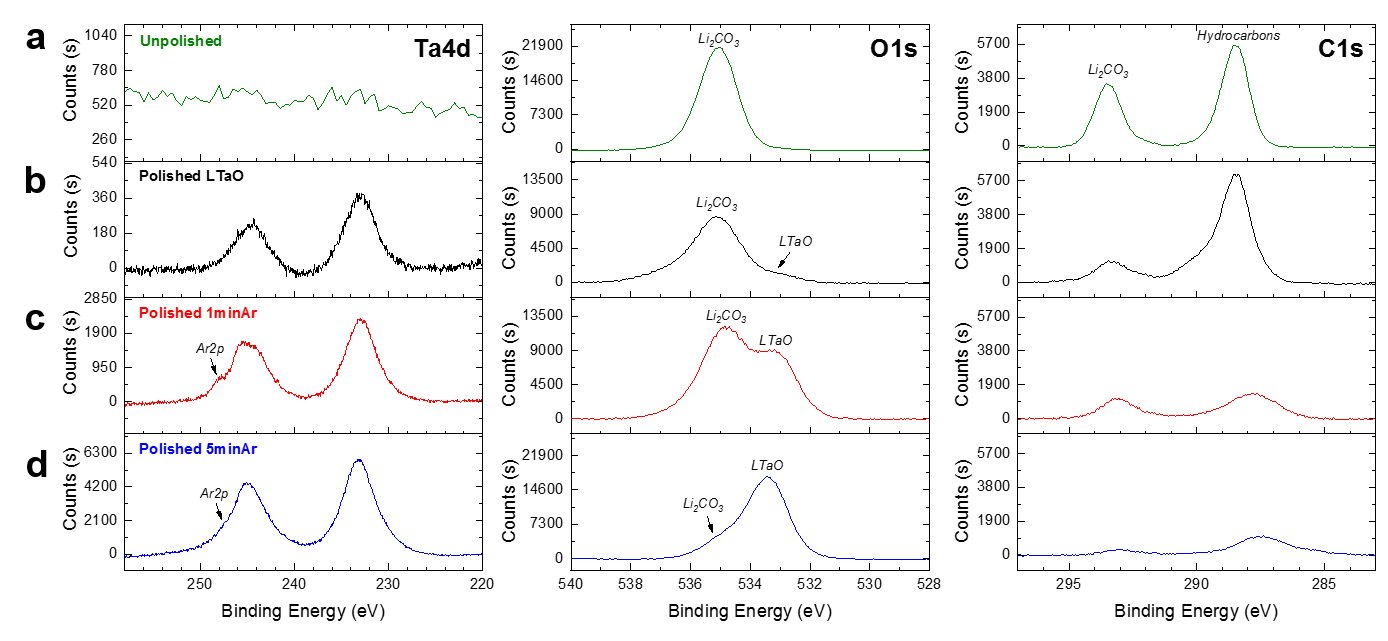}
    \caption{Ta4d, O1s and C1s XPS core level spectra recorded on \ce{Li7TaO6} pellets: (a) unpolished, (b) polished, (c) polished plus 1 min  Ar sputtering, and (d) polished plus 5 min Ar sputtering. 
}
    \label{fig:xps}
\end{figure*}

In \myfigref{fig:ion_cond} we show the Arrhenius plot of the ionic conductivities together with the fitted activation energies obtained for \ce{Li7TaO6}, xs-\ce{Li7TaO6}, \ce{Li_{7+x}Ta_{0.98}Zr_{0.02}O6}, and \ce{Li_{7+x}Ta_{0.98}Mo_{0.02}O6} prepared by pressing+sintering, and of the \ce{Li7TaO6} pellet prepared by the SPS method. 
For comparison, we also show the results from the simulations, together with the conductivity of LiPON~\cite{yu_stable_1997}.
We observe that the ionic conductivity of \ce{Li7TaO6} is two to three orders of magnitude higher than that of xs-\ce{Li7TaO6}.
This dramatic difference might be explained by the increased presence of side phases in the xs-\ce{Li7TaO6} pellet, as discussed in the context of pellet density and ICP-OES results above, or by the lower relative density of the xs-\ce{Li7TaO6} pellet itself, c.f. Table~\ref{tab:dens}. 
In line with this reasoning, the ionic conductivity of \ce{Li7TaO6} is significantly improved in the high-density pellet prepared by the SPS method. 
The substitution of Ta by 2\% Mo decreases the ionic conductivity in \ce{Li_{7+x}Ta_{0.98}Mo_{0.02}O6}, as well as the relative pellet density (c.f. Table~\ref{tab:dens}).
Vice versa, the \ce{Li_{7+x}Ta_{0.98}Zr_{0.02}O6} pellet exhibits a high ionic conductivity, on a par with the \ce{Li7TaO6} pellet prepared by SPS.
Both pellets display a high relative density, which confirms a close correlation between the relative pellet density and the ionic conductivity, as previously suggested~\cite{chen_influence_2002}. 
However, this correlation might be indirect rather than causal.
We propose the hypothesis that the amount of side phases present in the pellet influences both the relative density and the ionic conductivity, and thus represents their indirect link.

Even the highest experimentally measured ionic conductivities are still several orders of magnitude lower than the results from FPMD or PFF calculations, 
and the experimentally obtained activation energies at $\sim$0.6~eV are about two times those obtained from simulations.
The conductivity values reported in \myfigref{fig:ion_cond} correspond to the overall ionic conductivity of the sample, i.e. bulk plus grain boundaries. 
The Nyquist plot of the impedance of a polished \ce{Li7TaO6} pellet, shown in the inset of \myfigref{fig:ion_cond}, reveals two semi-circles that can be attributed to bulk resistance and grain-boundary resistance, respectively.
On the basis of the capacitance and dielectric constant, we can attribute the first semicircle to the bulk contribution ($C=10^{-11}$~F, $e=46$), while the second semicircle, with  eight times higher a capacitance, is attributed to grain-boundary contribution.
The last contribution, at low frequency, stems from  capacitance of the electrode. 

\begin{figure*}[t]
    \centering
    \includegraphics[width=\hsize]{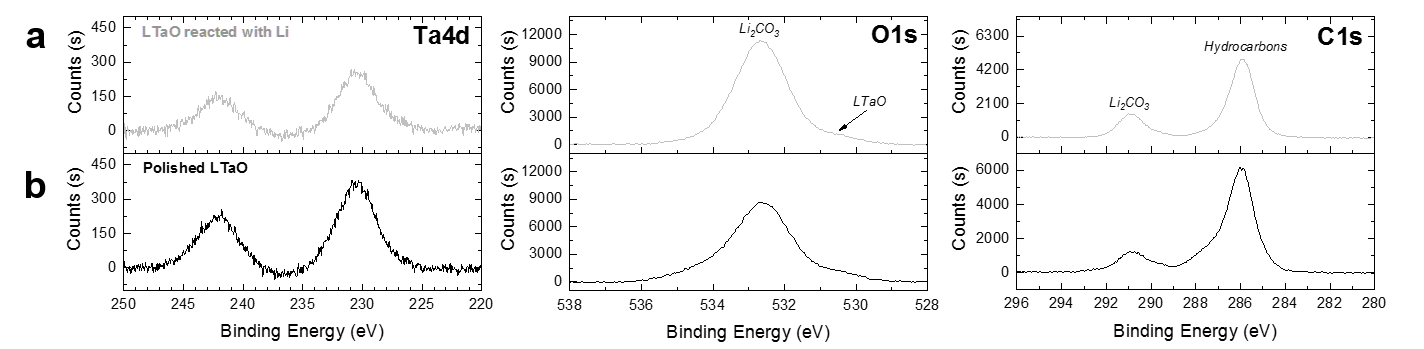}
    \caption{XPS spectra (Ta4d, O1s, C1s) for a polished \ce{Li7TaO6} pellet (a) before and (b) after exposure to metallic Li.}
    \label{fig:xps_stability}
\end{figure*}

We performed X-ray photoelectron spectroscopy (XPS) experiments in order to identify material characteristics such as surface reactivity and the presence of side phases.
Since \ce{Li7TaO6} is exposed to ambient air during the synthesis and pellet fabrication steps, we assume that the material reacts with \ce{CO2} and moisture to form surface species such as lithium carbonate (\ce{Li2CO3}) or lithium hydroxide (\ce{LiOH}). 
\myfigref{fig:xps} shows the XPS Ta4d, O1s, and C1s core level spectra of unpolished, polished (in an inert atmosphere), 
as well as Ar-sputtered (polished) \ce{Li7TaO6} pellets.
Unsurprisingly, the unpolished pellet surface showed no signal for Ta4d while a strong presence of \ce{Li2CO3} is revealed by the O1s and C1s spectra, depicted in \myfigref{fig:xps}a.
Since the XPS penetration depth is no more than 10~nm with an Al K$\alpha$ source, this \ce{Li2CO3} surface layer must be at least 10~nm thick.
However, the polished \ce{Li7TaO6} pellet exhibits a Ta4d signal as well as a shoulder (531 eV) in the O1s spectrum, which corresponds to exposed tantalate as the mother phase (see \myfigref{fig:xps}b).
Note that \ce{Li2CO3} and some hydrocarbons could still be detected in the XPS O1s and C1s spectra even after substantial polishing from a 2 mm thick pellet down to a thickness of $500-600$~$\mu$m, which prompted us to probe the pellet surface further by Ar sputtering.
Figs.~\ref{fig:xps}c and~\ref{fig:xps}d show the aforementioned core level spectra for polished pellets sputtered with Ar for 1~min and 5~min, respectively.
The \ce{Li2CO3} peaks in the O1s and C1s spectra substantially decreased while the signal of Ta increased in the Ta4d spectra, which indicates that \ce{Li2CO3} is mainly on the surface of the polished pellet.
We note that Ar sputtering for several minutes can alter the local composition of the material due to ion implantation, which leads to a shoulder (labeled Ar2p) in the Ta4d core level spectra shown in panels (c) and (d) of \myfigref{fig:xps}.
Nevertheless, regardless of polishing under inert atmosphere and Ar-sputtering, \ce{Li2CO3} contamination of the \ce{Li7TaO6} pellet could still be detected.

These experimental findings are in agreement with DFT results for the reaction energetics of \ce{Li7TaO6} with \ce{CO2}.
The \ce{Li7TaO6} phase itself was calculated to lie essentially on the convex hull of thermodynamic stability in the canonical Li-Ta-O phase diagram.
However, in presence of \ce{CO2} the Li carbonate formation was found energetically favoured by approximately $0.6$~eV per \ce{CO2} molecule at room temperature:
\begin{equation}
2\ \ce{Li7TaO6} \ + \ 7\ \ce{CO2} \quad \longrightarrow \quad \ce{Ta2O5} \ + \ 7\ \ce{Li2CO3},
\label{react:carbonate}
\end{equation}
where we included an entropic contribution $-TS=0.9$~eV to the free energy of gaseous \ce{CO2}, estimated from NIST reference data~\cite{NIST_Chem_WebBook} at a temperature $T=300$~K and pressure $p=4\times 10^{-4}$~bar (approx. atmospheric \ce{CO2} concentration).
Only above approximately $T$=500\,K, the left-hand side of reaction~(\ref{react:carbonate}) becomes entropically stabilized. 
\ce{Li2CO3} is a poor Li-ion conductor~\cite{mizusaki_lithium_1992}.
Therefore, the presence of poorly conductive side phases could contribute to the significant reduction of our experimentally measured ionic conductivity of \ce{Li7TaO6} pellets compared to the computated bulk ionic conductivity.
We conclude that exposure of the \ce{Li7TaO6} to \ce{CO2} must be carefully avoided at all stages of synthesis and characterization in order to prevent the formation of \ce{Li2CO3}, which imposes technical challenges to the laboratory environment.
Further improving the \ce{Li7TaO6} samples is therefore deferred to future work.
However, also the possibility of problems in the description of \ce{Li7TaO6} by DFT at the PBE level must be acknowledged.
We confirmed that the forces on the PBE level match perfectly the forces calculated with DFT+$U_{sc}$ (see \supplsecref{sec:sup-dftpU}), so self-interaction of the Ta-$d$ orbitals can be excluded as a source of error.
Another possibility is a phase transition to a superionic phase at high temperature, as hypothesized by M\"uhle \textit{et al.}~\cite{muhle_new_2004} who observed a drastic change in activation energy at 400~$^\circ$C, to 0.29~eV.
This barrier is the same as in the simulations we performed, which could indicate that our simulations sample a superionic phase also at ambient temperature.
However, we could not reproduce that change in activation barrier experimentally, therefore also this explanation remains speculative.

\subsection{Electronic conductivity}
The electronic conductivity of a material must be very low to allow for its application as SSE~\cite{goodenough_challenges_2010}.
Therefore, we measured the electronic conductivity of \ce{Li7TaO6} at room temperature, by applying a 100~mV DC voltage to a Cu/\ce{Li7TaO6}/Cu cell.
Due to the ion-blocking cell configuration, the steady state DC current reached after approximately 0.5~hours was attributed to the electronic conductivity of the \ce{Li7TaO6} solid-state electrolyte.
The corresponding value is $2 \times 10^{-9}$~S\,cm$^{-1}$, which is on a par with other SSE materials such as LLZO~\cite{rangasamy_role_2012} and LPS~\cite{minami_lithium_2007,shin_comparative_2014}.

\subsection{Electrochemical and chemical stability}
\label{subsec-electrochemical-results}

Our computational results yield a potential window of stable stoichiometry of \ce{Li7TaO6} between -0.85~V and 3.85~V vs. Li metal.
The negative lower potential limit means that \ce{Li7TaO6} cannot become Li super-stoichiometric upon contact with a metallic Li anode.
Vice versa, the large upper potential limit indicates good stability of the \ce{Li7TaO6} against oxidative Li extraction.
We estimate the potential window of phase stability of \ce{Li7TaO6} between 0.31~V and 2.51~V vs. Li metal, which is more limiting than the stable stoichiometry window.
The lower and upper potential limits are defined by the respective reductive and oxidative decomposition reactions:

\begin{align}
& \ce{Li7TaO6} \ + \ 5~\ce{Li} \quad \longrightarrow \quad \ce{Ta} \ + \ 6~\ce{Li2O} \\
& \ce{Li7TaO6} \quad \longrightarrow \quad \ce{Li3TaO4} \ + \ \ce{O2} \ + \ 4~\ce{Li},
\end{align}
where we include an entropic contribution of $-TS =-0.638$~eV to the free energy of gaseous \ce{O2} at a temperature of $T=300$~K and pressure of $p=1$~bar, which was computed from NIST reference data~\cite{NIST_Chem_WebBook}.
The lower stability potential limit of 0.31~V is only slightly positive vs. Li metal, and thus \ce{Li7TaO6} is potentially metastable in contact with a metallic Li anode.
Compared to other common Li-SSE materials, the \ce{Li7TaO6} stability window is excellent and close to the one of LLZO~\cite{zhu_origin_2015}, one of the most stable SSE materials known to date.

The reactivity of \ce{Li7TaO6} against metallic Li was experimentally elucidated by placing a polished \ce{Li7TaO6} pellet in contact with Li foil under pressure in a Swagelok cell. To promote the reaction between said materials, the cell was exposed to a high temperature (80$^{\circ}$C) as described in \mysecref{subsec-experimental}.
\myfigref{fig:xps_stability} shows the XPS Ta4d, O1s, and C1s core level spectra of the \ce{Li7TaO6} pellet surface before and after exposure to metallic Li. 
We observe no significant changes in the XPS spectra.
Specifically, the oxidation state of Ta does not change after exposure to Li metal even at elevated temperatures, which indicates stability, or at least meta-stability, against metallic Li, in good agreement with the computational results.

\begin{figure}[t]
    \centering
    \includegraphics[width=\hsize]{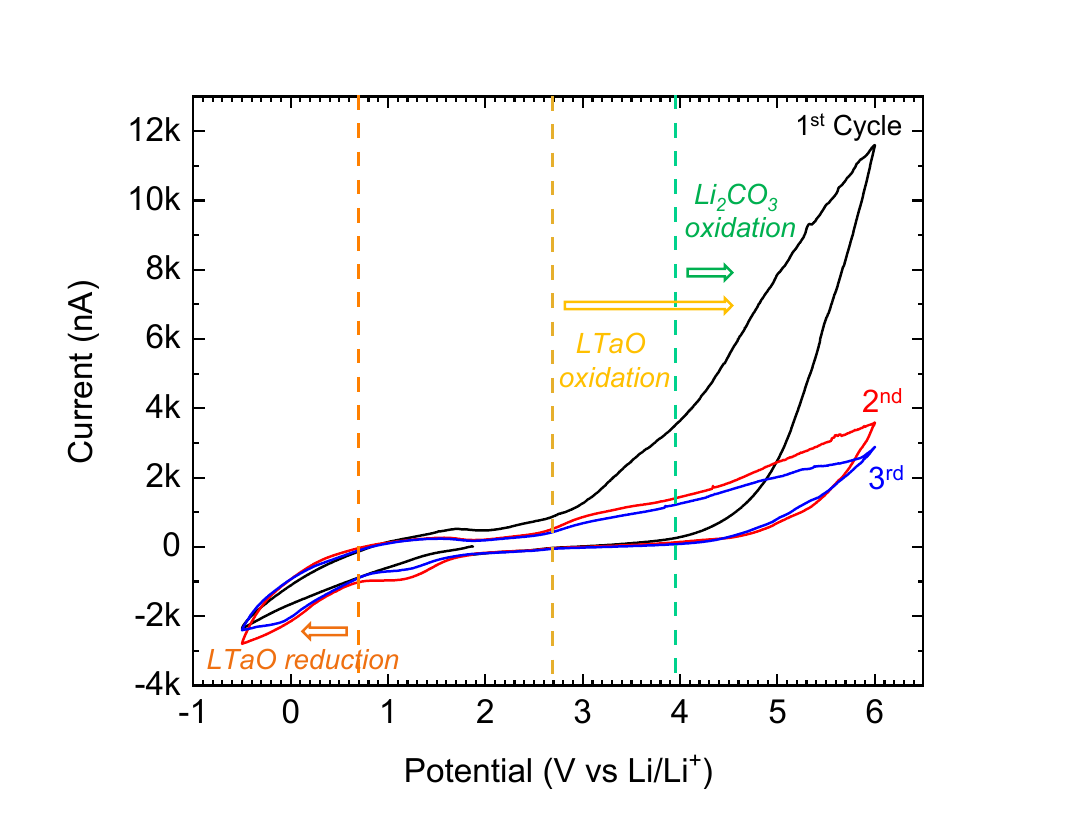}
    \caption{CV curves of a Au-coated \ce{Li7TaO6} pellet with Li as the reference and counter electrode. The measurements were conducted at 80~$^{\circ}$C at a scan rate of 0.1~mV/s.}
    \label{fig:cv}
\end{figure}

The electrochemical stability window of \ce{Li7TaO6} was evaluated using cyclic voltammetry (CV) with a semiblocking cell configuration (Au-\ce{Li7TaO6}-Au/Li).
The resulting cyclic voltammograms shown in \myfigref{fig:cv} were taken over a wide voltage range (up to 6~V) at a slow scan rate of 0.1~mV\,s$^{-1}$.
In the first cycle, oxidation processes are apparent above $\sim$2.7~V vs. Li.
However, we observe no corresponding reduction processes. 
Since the presence of \ce{Li2CO3} in the \ce{Li7TaO6} pellet is confirmed by XPS measurements, we believe that lithium carbonate decomposition is responsible for the irreversible oxidation reaction at higher potentials in the initial cycle.
Mahne \textit{et al.}~\cite{mahne_electrochemical_2018} reported that \ce{Li2CO3} decomposes into carbon dioxide above 3.8~V vs. Li. 
In the second and third cycles, the irreversible oxidation current substantially decreased at higher potentials, while a smaller oxidation shoulder persisted around 2.7~V vs. Li, which is in very good agreement with the computed 2.51~V upper stability potential limit of \ce{Li7TaO6}. 
Vice versa, a reduction process is observed with an onset potential at $\sim$0.7~V vs. Li, which is close to the computed  lower stability potential limit at 0.31~V.  There is a second reduction process with an onset around 1.7 V~vs. Li that we cannot clearly identify.
It could correspond to the reduction either of oxidation products previously formed at high potentials or of synthesis residuals.\section{Conclusions}
\label{sec:conclusions}

We studied the hexa-oxometallate \ce{Li7TaO6} using accurate first-principles  as well as polarizable force-field molecular dynamics simulations, finding excellent agreement between the two methods. 
From our simulations we conclude that the compound investigated is a fast-ionic conductor also at room temperature.
The activation barriers of 0.27~eV and 0.29~eV, estimated from the FPMD and PFF simulations, respectively, are of sufficiently low value to highlight this material as a potential candidate material for SSE application.
We could not confirm the high ionic conductivity and low barrier with experiments:
Our experimental results indicate that carefully avoiding the formation of side phases at all stages of synthesis and experiments, although technically very challenging, will be required to obtain better agreement between experimental and theoretical conductivities.
In addition, we calculated the electrochemical stability window from first principles, finding that \ce{Li7TaO6} is remarkably stable, and is comparable to LLZO in this respect.
The computational findings are in good agreement with the results of electrochemical stability experiments, namely cyclic voltammetry and direct reactivity with metallic Li.

\section*{Acknowledgements}
We all thank T. Laino for his extensive help in liaising different teams and many fruitful discussions.
We also thank R. Haumont and C. Byl from SP2M/ICMMO/Universit\'e Paris-Saclay and the Plateforme de Frittage \^Ile de France (Thiais, France) for their help with the SPS synthesis.
L.K. thanks I. Timrov for helpful discussions and advice on the DFT+$U_{sc}$ calculations.
We gratefully acknowledge support from the Swiss National Science Foundation (MARVEL NCCR and project 200021-159198), and the Swiss National Supercomputing Centre CSCS (project s836).

\section*{Author contributions}
L.K., N.M., and D.P. conceived the project.
L.K. conducted and analyzed the FPMD simulations.
T.B. and A.M. performed the computational electrochemical stability analysis.
X.C. and E.G. synthesized all samples and performed the experimental characterization (XRD, ionic conductivity).
F.Z. fitted the parameters for the PFF and conducted the large PFF simulations.
A.M. and L.K. also ran PFF simulations and analyzed the results.
S.D.L. performed the electrochemical tests, and S.D.L. and M.E.K. supported the XPS characterization.
All authors contributed to the writing and discussion of the manuscript.

\appendix

\FloatBarrier
\bibliographystyle{bibliostyle}
\bibliography{bibliography}
%
\balance

\end{document}